\documentstyle[epsf]{l-aa}

\def\lta{\hbox{\raise.5ex\hbox{$<$}
\kern-1.1em\lower.5ex\hbox{$\sim$}}}
\def\gta{\hbox{\raise.5ex\hbox{$>$}
\kern-1.1em\lower.5ex\hbox{$\sim$}}}

\newcommand{\aaa}{A\&A}      
\newcommand{\apj}{ApJ}       
\newcommand{\mnras}{MNRAS}   
\newcommand{\betaac}{\beta^{2}_{a_c}}

\renewcommand{\Im}{\rm Im}
\begin{document}

\thesaurus{11(11.10.1; 02.08.1; 02.09.1)}

\title{Kelvin-Helmholtz instability of stratified jets}

\author{M. Hanasz\inst{1} \and H. Sol\inst{2}}

\offprints{M. Hanasz}

\institute{Institute of Astronomy, Nicolaus Copernicus University,
 ul. Chopina 12/18, PL-87-100 Torun, Poland.
\and
DARC, UPR176 du CNRS, Observatoire de Paris-Meudon, 5 Place J. Janssen,
F-92195 Meudon Cedex, France.}

\date{Received January 24/ accepted March 13, 1996}

\maketitle

\begin{abstract}
We investigate the Kelvin-Helmholtz instability of stratified jets. The
internal component (core) is made of a relativistic gas moving with a
relativistic bulk speed. The second component (sheath or envelope) flows
between the core and external gas with a nonrelativistic speed.  Such a
two-component jet describes a variety of possible astrophysical jet
configurations like e.g. (1) a relativistic electron-positron beam
penetrating a classical electron-proton disc wind or (2) a beam-cocoon
structure. We perform a linear stability analysis of such a configuration
in the hydrodynamic, plane-parallel, vortex-sheet approximation.  The
obtained solutions of the dispersion relation show very apparent
differences with respect to the single-jet solutions.

Due to the reflection of sound waves at the boundary between sheet and external
gas, the growth rate as a function of wavenumber presents a specific
oscillation pattern. Overdense sheets can slow down the growth rate and
contribute to stabilize the configuration. Moreover, we obtain the result
that even for relatively small sheet widths the properties of sheet start
to dominate the jet dynamics. Such effects could have important astrophysical
implications, for instance on the origin of the dichotomy between radio-loud
and radio-quiet objects.

\keywords{Galaxies: jets -- Hydrodynamics -- Instabilities}

\end{abstract}

\section{Introduction}

Kelvin-Helmholtz instability due to shearing induced by large velocity
gradients is well known to develop in extragalactic jets. This has been
shown by several analytical and numerical works such as the ones reviewed
for instance by Birkinshaw, 1991. These studies often consider the
jet as a single fluid with one bulk velocity and a single interface made by
a sheared layer with the external medium.  However there exist both
observational and theoretical pieces of evidence supporting the fact
that components with different bulk velocities are present inside jets
themselves.  Among the observational data showing the presence of
different velocities inside jets we can mention (i) the detection in
some compact sources of VLBI components with different apparent speed,
(ii) the superluminal effect which proves that some jets are highly
relativistic at VLBI scale while properties of the extended component
rather suggest classical or mildly relativistic large scale jets, (iii)
the presence in some sources of entrained gas which emits optical lines
from which one can infer velocities likely smaller than the main bulk
velocity of the jet.  The peculiar morphology of the wide angle tail radio
sources (WATs) appears also very difficult to explain (Eilek et al, 1984;
O'Donoghue et al, 1990 and 1993) except if one assumes that matter with
two different speeds is present in the jet, namely one relativistic
component radiating before the inner hot spot and one component with
slow speed radiating after the inner hot spot (Leahy, 1984).
Recent data on M87 provide the first direct measurement of apparent bulk
velocity at the kiloparsec scale and show a complex velocity pattern
with the presence of quite different bulk flow velocities along the jet
(Biretta et al, 1995). In 3C273, the radio emitting jet appears longer
and wider than the optical one. Bahcall et al (1995) suggest that there
are actually two superposed components, a fast-moving inner jet
surrounded by a slow-moving ``cocoon''.  Besides that, a Fanaroff-Riley
type II radio source can also be considered as a stratified jet with two
different layers corresponding to the inner real jet and to the backflow
forming the surrounding cocoon.

A quite new way of looking at the radio data appears as well in favour
of the existence of several components in radio sources. Starting with the
use of color-color diagrams (Katz-Stone et al, 1993), Rudnick et al, 1994
and Katz-Stone and Rudnick (1994) show that it is reasonable to assume
a distribution of radiating particles which is not a power-law throughout
the sources and can then partially isolate the
different parameters which contribute to the synchrotron brightness,
namely the magnetic field B, the number of radiating particles $N_T$ and
some fiducial Lorentz factor $\gamma_0$ that characterizes their energy
distribution. Adding some knowledge concerning the shape of the radiating
particle distribution, they can produce frequency-independent maps,
proportional to $N_TB$, $\gamma^2_0B$ and $N_T/\gamma_0^2$. This powerful
method provides a completely changed view of the sources in which new
features are discovered. For instance, in the eastern lobe of Cygnus A,
they detect an edge-brightened channel girdled by rings, which corresponds
to a real enhancement of the radiating particle density, likely located
around the counterjet (Katz-Stone, Rudnick, 1994). Recent developments of
their technics lead them to suggest that jets in both FRI and FRII
sources may have coaxial sheaths a few times wider than the jets
themselves. From their ``tomography'' analysis of radio data which
combine maps at different frequencies, they detect two-component
structures in 3C449 and 1231+674, with flat spectrum jets surrounded by
sheaths of steep spectrum emission with different polarization properties
(Rudnick, 1995). It is not yet known whether such sheaths emanate from
the inner jets or have been directly ejected from the central engines.
However their presence deeply emphasizes the likelihood of
multi-component jet models.

 Indeed several models of jet formation and propagation reach the concept
of stratified jets.  First Chan and Henriksen, 1980, mentioned the
``multilevel'' structure as a basic modification to add to their
self-similar jet model in order to explain why jets appear to have nozzles
on very different scales. Smith and Raine, 1985, investigate a two-level
configuration and explore the possibility to produce collimated outflows
by the interaction of a nuclear wind from the very inner region of an
active galactic nucleus with a Compton-heating-induced wind from an
accretion disc. Baker et al, 1988, deal with another kind of two-component
model while studying the radiative properties of a relativistic particle
beam injected into an extragalactic jet and the possibility of collective
emission of radio waves. A completely different approach by Melia and
K\"onigl, 1989, studies the Compton-drag deceleration of ultrarelativistic
nuclear jets.  It predicts the existence of a transverse gradient in the
asymptotic bulk Lorentz factor distribution of the particles in the
radio jets. K\"onigl and Kartje (1994) also consider highly stratified
winds from accretion discs, with inner ionized gas and outer dusty neutral
outflow. Sol, Pelletier, Ass\'eo,
1989, propose a two-component model taking into account a relativistic
electron-positron beam extracted from the funnel of an accretion disc and
streaming through a classical electron-proton collimated wind coming out
from all parts of the disc.  This model provides a simple explanation to
the velocity dilemma if VLBI jets and superluminal motion are related to
the relativistic beam while kiloparsec scale radio features are associated
to the slower collimated wind.  Stability of such a configuration
relatively to the excitation of Langmuir, Alfven and whistler waves has
been found for electron-positron beams propagating along strong enough
magnetic field with small enough Lorentz factor (Sol et al, 1989;
Achatz et al, 1990; Pelletier and Sol, 1992;  Achatz and Schlickeiser, 1992).
The condition for stability against excitation of Langmuir waves
requires a strong longitudinal magnetic field $B$ such that the electron
gyrofrequency is larger than the plasma frequency in the wind, namely
$B>3.2\times 10^{-3} \sqrt{n_p}$ in CGS units where $n_p$ is the wind
density.  The constraint on the Lorentz factor $\gamma$ comes from the
necessity to avoid excitation of Alfven waves which imposes $\gamma <
\sqrt{m_p\over m_e} \simeq 43$.  Therefore it appears possible to quench
the beam-plasma instability and to ensure a two-component configuration
stable from the point of view of microphysics.  However, the question of
large scale fluid instability of such two-component jet is still open.
It is investigated in the present work. As a first approach to the study
of the Kelvin-Helmholtz stability of stratified jets, we consider here
a core-sheet structure. It is the simplest ``multilevel'' configuration
which appears in some jet models (Smith and Raine, 1985;  Baker et al,
1988;  Sol et al, 1989) and provides an approximation to the
general case of jets surrounded by cocoons or sheaths.

\begin{figure}     
\epsfxsize=\hsize \epsfbox{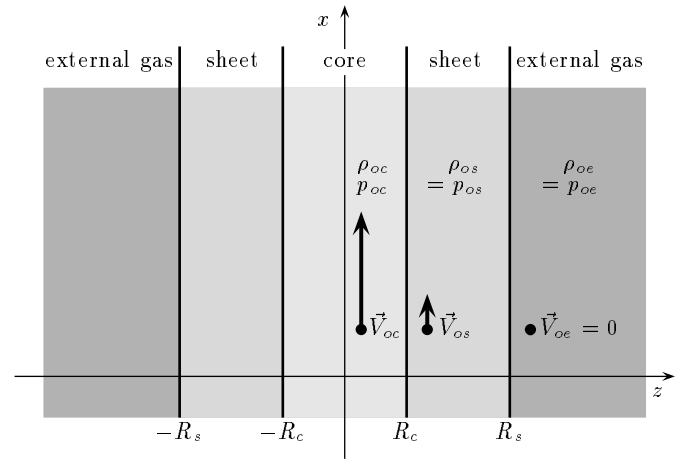}
\caption[]{
Schema of a stratified jet with core, sheet and external
components in the unperturbed state.
}
\end{figure}

\section{The dispersion relation}

We describe the jet with the 2-dimensional slab geometry, which provides
an adequate description of the more realistic cylindrical case with respect
to the Kelvin-Helmholtz instability (Ferrari et al. 1982). For a
large range of physical parameters, the solutions for both the slab and
cylindrical cases have similar behaviour, apart from slight numerical
differences.  This similarity associates the symmetrical and antisymmetrical
perturbations of the slab jet with the pinching and helical perturbations of
the cylindrical jet respectively.  Only high order fluting modes do not have
counterparts in slab jets (Ferrari et al, 1982; Hardee, Norman, 1988).

Both spatial and temporal stability
analysis of Kelvin-Helmholtz instability have been carried out in the
literature. Spatial and convected temporal growth rates have been shown
to be not equal but still similar in the two approaches (Hardee, 1986;
Norman, Hardee, 1988). This paper investigates the temporal behavior of
the instability.  We consider a core-sheet jet made of
three layers as shown in  Fig.~1 and describe the transition layers at
all interfaces in the vortex-sheet approximation.

The 2-dimensional core-sheet jet initially flows along the $x$-axis
on the $x-z$ plane.  The unperturbed equilibrium state is defined as
follows.  $R_c$ corresponds to the radius of the inner core, $R_s$ to
the radius of the surrounding sheet.  $V_{o_c}$ $(V_{o_s}, V_{o_e})$ is
the bulk velocity in the core (respectively the sheet and the external
medium), $a_c$ the sound velocity in the core, $\rho_c$ $(\rho_s$,
$\rho_e)$ and $p_c$ $(p_s$, $p_e)$ the proper density and pressure in the
core (respectively the sheet and the external medium). We also introduce
the density contrasts $\nu_s = {\rho_{o_c}/ \rho_{o_s}}$ and $\nu_e =
{\rho_{o_s}/ \rho_{o_e}}$, where index $o$ stands for unperturbed zero-order
values.  The relativistic Lorentz factor of the core assumed to be a
relativistic gas, the Mach number of the core and the Mach number of sheet
are respectively
\begin{eqnarray}
\gamma &=&\left(1-\frac{V_{o_c}^2}{c^2}\right)^{-1/2}, \\
M_c &=& \frac{V_{o_c}}{a_c},\\
M_s &=& \frac{V_{o_s}}{a_s}.
\end{eqnarray}
We assume that the gas of the sheet moves with a
non-relativistic bulk velocity, $V_{o_s} \ll c$, and is described by a non
relativistic equation of state with adiabatic index $\Gamma_s = 5/3$, however
mildly relativistic sheet could also be taken into account within an
analogous way. The external gas is also represented by a nonrelativistic
gas with adiabatic index $\Gamma_e =5/3$, with $V_e=0$ in the unperturbed
state.

We shall use the relativistic equation of hydrodynamics in the form applied by
Ferrari et al. (1978) for the relativistic core

\begin{equation}
 \gamma^2 \!\!\left( \rho_c \!+\! {p_c\over c^2} \right)\!\! \left[ {\partial
   \vec V_c\over \partial t}\!+\! (\vec V_c \cdotp \vec\nabla) \vec V_c \right]
  + \vec\nabla p_c + {\vec V_c\over c^2} {\partial p_c \over \partial t}
  = 0 \label{eq:1}
\end{equation}
\begin{equation}
 \gamma \left( {\partial \rho_c\over \partial t} \!+\! \vec V_c
  \vec \nabla\rho_c \right) \!+\! \left( \rho_c + {p_c\over c^2} \right)\!\!
 \left[ {\partial \gamma \over \partial t} + \vec\nabla (\gamma \vec V_c)
  \right] = 0  \label{eq:2}
\end{equation}

\noindent and for the non-relativistic sheet gas and external medium,

\begin{equation}
 {\partial \vec V_{s,e}\over \partial t} + (\vec V_{s,e} \vec\nabla)
 \vec V_{s,e} + {\vec\nabla p_{s,e}\over \rho_{s,e}} = 0 \label{eq:3}
\end{equation}
\begin{equation}
 {\partial \rho_{s,e}\over \partial t} + \vec\nabla (\rho_{s,e}\,
  \vec V_{s,e}) = 0 \ . \label{eq:4}
\end{equation}

The rest frame density of the relativistic core gas  and the enthalpy are
respectively
\begin{eqnarray}
\rho_c &=& m_c n_c + \frac{1}{\Gamma_c-1}\frac{p_c}{c^2},\\
w_c &=& \rho_c  + \frac{p_c}{c^2}
\end{eqnarray}
where $m_c$ and $n_c$ are are the rest mass and the number density of gas.
The sound speed of relativistic gas is (Taub 1948, Landau \& Lifshitz, 1959)
\begin{equation}
a_c^2 = \frac{\Gamma_c p_c}{w_c}
\end{equation}
In the ultrarelativistic limit the adiabatic index $\Gamma_c = {4/ 3}$, then
\begin{equation}
a_c = (\Gamma_c-1)^{1/2} c = \frac{c}{\sqrt{3}}.
\end{equation}
The Lorentz factor of the core expressed by dimensionless quantities is
\begin{equation}
\gamma = (1-\betaac\, M_c^2)^{-1/2},
\end{equation}
where $\betaac={a_c^2/ c^2} = {1/ 3}$.

For the nonrelativistic sheet and external gases we consider equations of
state
\begin{equation}
 p_{s,e} \rho_{s,e}^{-\Gamma_{s,e}} = p_{o_{s,e}}
   \,\rho_{o_{s,e}}^{-\Gamma_{s,e}} =\ \hbox{constant}\ , \label{eq:5}
\end{equation}
with $\Gamma_{s,e} = 5/3$, hence
\begin{equation}
  \frac{\rho_{s,e}}{\rho_{o_{s,e}}} = \left( \frac{p_{s,e}}{p_{o_{s,e}}}
  \right)^{1/\Gamma_{s,e}} \ .
\end{equation}
The sound speeds of the sheet and external gases are
\begin{equation}
a_{s,e} =\left( \frac{\Gamma_{s,e} p_{s,e}}{\rho_{s,e}}\right)^{1/2}
\end{equation}
and the density contrasts can be expressed by sound speed ratios
\begin{eqnarray}
\nu_s &=& {\rho_{o_c} \over \rho_{o_s}} = {{a_s}^2 \over
        \Gamma_{s}\, a^2_c},\\
\nu_e &=& \frac{\Gamma_s a_e^2}{\Gamma_e a_s^2} \ .
\end{eqnarray}

 {}From now on we shall use dimensionless quantities with sizes scaled to
$R_c$ and time scaled to ${ R_c/ a_{o_c}}$ and introduce the
relative pressure perturbations
\begin{equation}
g_{c,s,e} = {p_{c,s,e}/ p_o}
\end{equation}
where $p_o =p_{o_c} = p_{o_s} = p_{o_e}$.

The equations describing the
relativistic core gas can be written as
\begin{equation}
 \Gamma_c \gamma^2 g_c \left[ {\partial \vec V_c\over \partial t} +
 (\vec V_c \vec \nabla) \vec V_c \right] + \vec\nabla g_{c} + \betaac\,
 \vec V_c {\partial g_c\over \partial t} = 0 \label{eq:6}
\end{equation}
\begin{equation}
 \gamma \left[ {\partial g_c \over \partial t} + \vec V_c \vec\nabla g_c
  \right] + \Gamma_c \,g_c \left[ {\partial\gamma \over \partial t} +
 \nabla (\gamma \vec V_c) \right] = 0 \ . \label{eq:7}
\end{equation}
and the equations of motion for the sheet and the external medium as
\begin{equation}
 {\partial \vec V_s\over \partial t} + (\vec V_s \vec\nabla)
   \vec V_s + \nu_s\, g_s^{-{1\over \Gamma_s}}
     \vec\nabla g_s = 0 \label{eq:8}
\end{equation}
\begin{equation}
 {\partial g_s^{1\over \Gamma_s}\over \partial t} + \vec\nabla
 \left( g_s^{1\over \Gamma_s}\, \vec V_s \right) = 0 \label{eq:9}
\end{equation}
\begin{equation}
 {\partial \vec V_e\over \partial t} + (\vec V_e \vec\nabla)
   \vec V_e + \nu_s \nu_e\, g_e^{-{1\over \Gamma_e}}
     \vec\nabla g_e = 0 \label{eq:8a}
\end{equation}
\begin{equation}
 {\partial g_e^{1\over \Gamma_e}\over \partial t} + \vec\nabla
 \left( g_e^{1\over \Gamma_e}\, \vec V_e \right) = 0 \label{eq:9a}
\end{equation}
The linearization of (\ref{eq:6}) and (\ref{eq:8}) leads to
\begin{equation}
 \gamma^2 \Gamma_c \left( {\partial\over \partial t}
    + M_c {\partial\over \partial x}\right) V'_{c_z} + {\partial g'_c\over
 \partial z} = 0 \label{eq:10}
\end{equation}
\begin{equation}
 \gamma^2 \Gamma_c \left( {\partial\over \partial t}
    + M_c {\partial\over \partial x}\right) V'_{c_x} + \left(
  {\partial \over \partial x} + \betaac\, M_c {\partial\over \partial t}
 \right) g'_c = 0 \label{eq:11}
\end{equation}
\begin{equation}
 {1\over \nu_s} \left( {\partial\over \partial t} + M_s
\sqrt{\nu_s\Gamma_s}{\partial\over
     \partial x} \right) V'_{s_z} + {\partial g'_s \over \partial z}
    = 0 \label{eq:12}
\end{equation}
\begin{equation}
 {1\over \nu_s\nu_e} \left( {\partial\over \partial t} + M_s
\sqrt{\nu_s\Gamma_s}{\partial\over
     \partial x} \right) V'_{s_x} + {\partial g'_s \over \partial x}
    = 0 \label{eq:13}
\end{equation}
\begin{equation}
 {1\over \nu_s\nu_e} {\partial V'_{e_z}\over \partial t}
    + {\partial g'_e\over \partial z}  = 0 \label{eq:14}
\end{equation}
\begin{equation}
 {1\over \nu_s\nu_e} {\partial V'_{e_x}\over \partial t}
    + {\partial g'_e\over \partial x}  = 0 \label{eq:15}
\end{equation}
in the rest frame of the external medium, where prime denotes first order
perturbed quantities. Combining (\ref{eq:10}), (\ref{eq:11}) and linearized
(\ref{eq:7}) gives
\begin{eqnarray}
\left.{\partial^2 g'_c\over \partial z^2} + \gamma^2 \right[&-& \left(
 {\partial\over \partial t} + M_c {\partial\over \partial x}\right)^2
\nonumber \\ &+& \left( {\partial\over \partial x}
 +\left. \betaac\, M_c {\partial\over \partial t}
 \right)^2 \right] g'_c = 0 \label{eq:16}
\end{eqnarray}
while combining linearized (\ref{eq:9}) with (\ref{eq:12}) and (\ref{eq:13})
for the sheet and with (\ref{eq:14}) and (\ref{eq:15}) for the external
medium gives
\begin{equation}
 {\partial^2 g'_s\over \partial z^2} +\left[ {\partial^2\over\partial x^2}
  - {1\over \nu_s \Gamma_s} \left( {\partial\over \partial t} +
M_s\sqrt{\nu_s\Gamma_s}
 {\partial \over \partial x}\right)^2 \right] g'_s = 0 \label{eq:17}
\end{equation}
and
\begin{equation}
 {\partial^2 g'_e\over \partial z^2} +\left[ {\partial^2\over\partial x^2}
 - {1\over \nu_s\nu_e \Gamma_e} {\partial^2\over \partial t^2}\right] g'_e = 0
  \ .  \label{eq:18}
\end{equation}

Let us assume in the core and the sheet perturbations of the form (for
$z> 0$)
\begin{equation}
\delta '_{c,s} = [ \delta^+_{c,s} F^+_{c,s} (z)
    + \delta^-_{c,s} F^-_{c,s} (z) ] \exp i (k_x x - \omega t) + c.c.
\end{equation}
where $F^{\pm}_{c,s} (z) = \exp (\pm i\, k_{{c_z,s_z}} z)$ allow to describe
waves propagating in opposite $z$-directions.  Perturbations in the
external medium are of the form
\begin{equation}
\delta '_e = \delta^+_e F^+_e (z) \exp\, i (k_x x-\omega t) +c.c.
\end{equation}
where $F^+_e (z) = \exp \, i\,
k_{e_z} z$ represents only outgoing waves for positive $z$.  Here $k_x$
is the parallel wavenumber longitudinal to the jet and $k_{c_z}$
(respectively $k_{s_z}$ and $k_{e_z}$) is the transverse wavenumber
perpendicular to the jet axis in the core region (respectively in the
sheet and external regions).  The transverse wavenumbers
$k_{c_z}$, $k_{s_z}$ and $k_{e_z}$ are complex since the perturbations we
consider represent surface waves.

For a temporal stability analysis, $k_x$ is
real and $\omega$ complex.  Equations (\ref{eq:16}), (\ref{eq:17}) and
(\ref{eq:18}) lead to
\begin{eqnarray}
 k_{c_z} &=& \left( \omega^2_{c_o} - k^2_{x_o} \right)^{1/2} \label{eq:19}\\
 k_{s_z} &=& \left( {\omega^2_{s_o} \over \nu_s \Gamma_s}
    - k_x^2 \right)^{1/2} \label{eq:20}\\
 k_{e_z} &=& \left( {\omega^2 \over \nu_s\nu_e \Gamma_e}
    - k_x^2 \right)^{1/2} \label{eq:21}
\end{eqnarray}
where $\omega_{c_o} = \gamma (\omega - M_c k_x)$ and $k_{x_o} = \gamma
(k_x - \betaac M_c \omega)$ are the frequency and wavenumber in the rest
frame of the core fluid and $\omega_{s_o} = \omega -\sqrt{\nu_s\Gamma_s}M_s k_x $ is the
frequency in the rest frame of the sheet fluid.  From equations (10),
(\ref{eq:12}) and (\ref{eq:14}) we deduce the relations
\begin{eqnarray}
 V'^\pm_{c_z} &=& {\pm k_{c_z} \over \Gamma_c \gamma\, \omega_{c_o}} \,
    G'^\pm_c \label{eq:22}\\
 V'^\pm_{s_z} &=& {\pm \nu_s k_{s_z}\over \omega_{s_o}}\, G'^\pm_s
  \label{eq23} \\
 V'^+_{e_z} &=& {\nu_s\nu_e k_{e_z}\over \omega}\, G'^+ _e  \label{eq:24}
\end{eqnarray}
Symmetry conditions of the problem allow to consider only half-space
with $z\geq 0$. The components $(+)$ and $(-)$ propagate respectively in the
positive and negative $z$-direction.  Our perturbation takes into
account only outgoing solution in the external medium.

 The dispersion relation of the system is obtained from the boundary
conditions at the internal (core/sheet) and external (sheet/external gas)
interfaces, namely pressure equilibrium and equality between the
$z$-component of the gas displacement and the transversal displacement of
contact surface.  In the reference frame of the core gas, this condition
writes
\begin{equation}
 v^{(c)'}_{c_z} - {\partial\over \partial t^{(c)}} h^{(c)'}_c = 0
  \label{eq:25}
\end{equation}
We call $h'_c$ and $h'_s$ the transversal displacements of core/sheet
and sheet/external surfaces. The exponent $(c)$ stands for values
determined in the core gas rest frame. From Lorentz transformation
\begin{equation}
v^{(c)'}_{c_z} = \frac{v_{c_z}' \sqrt{1-\betaac M_c^2}}{ 1-\betaac (M_c
     +v'_{c_x}) M_c} \ ,
\end{equation}
which reduces to $v^{(c)'}_{c_z} = \gamma v_{c_z}'$ in the linear
approximation. Moreover

\begin{equation}
{\partial\over \partial t^{(c)}} = \gamma \left( {\partial\over
\partial t} + M_c {\partial\over \partial x} \right)
\end{equation}
and $h'_c$, the displacement of surface, is Lorentz invariant as well as
the pressure $g'_c$. Boundary conditions at the internal interface
located at $z=1$ become
\begin{eqnarray}
 v'_{c_z} &-& \left( {\partial\over \partial t} + M_c {\partial\over
    \partial x} \right) h'_c = 0  \label{eq:26}\\
 v'_{s_z} &-& \left( {\partial\over \partial t} + \sqrt{\nu_s\Gamma_s}M_s {\partial\over
    \partial x} \right) h'_c = 0  \label{eq:27} \\
 g'_c &=& g'_s \label{eq:28}
\end{eqnarray}
and at the external interface located at $z=R$ $(\equiv {R_s/ R_c})$,

\begin{eqnarray}
 v'_{s_z}&-& \left( {\partial\over \partial t} + \sqrt{\nu_s\Gamma_s}M_s {\partial\over
    \partial x} \right) h'_s = 0  \label{eq:29} \\
 v'_{e_z}&-& {\partial\over \partial t}  h'_s = 0  \label{eq:30} \\
 g'_s&=& g'_e\ . \label{eq:31}
\end{eqnarray}

\begin{figure*}     
\epsfxsize=\hsize \epsfbox{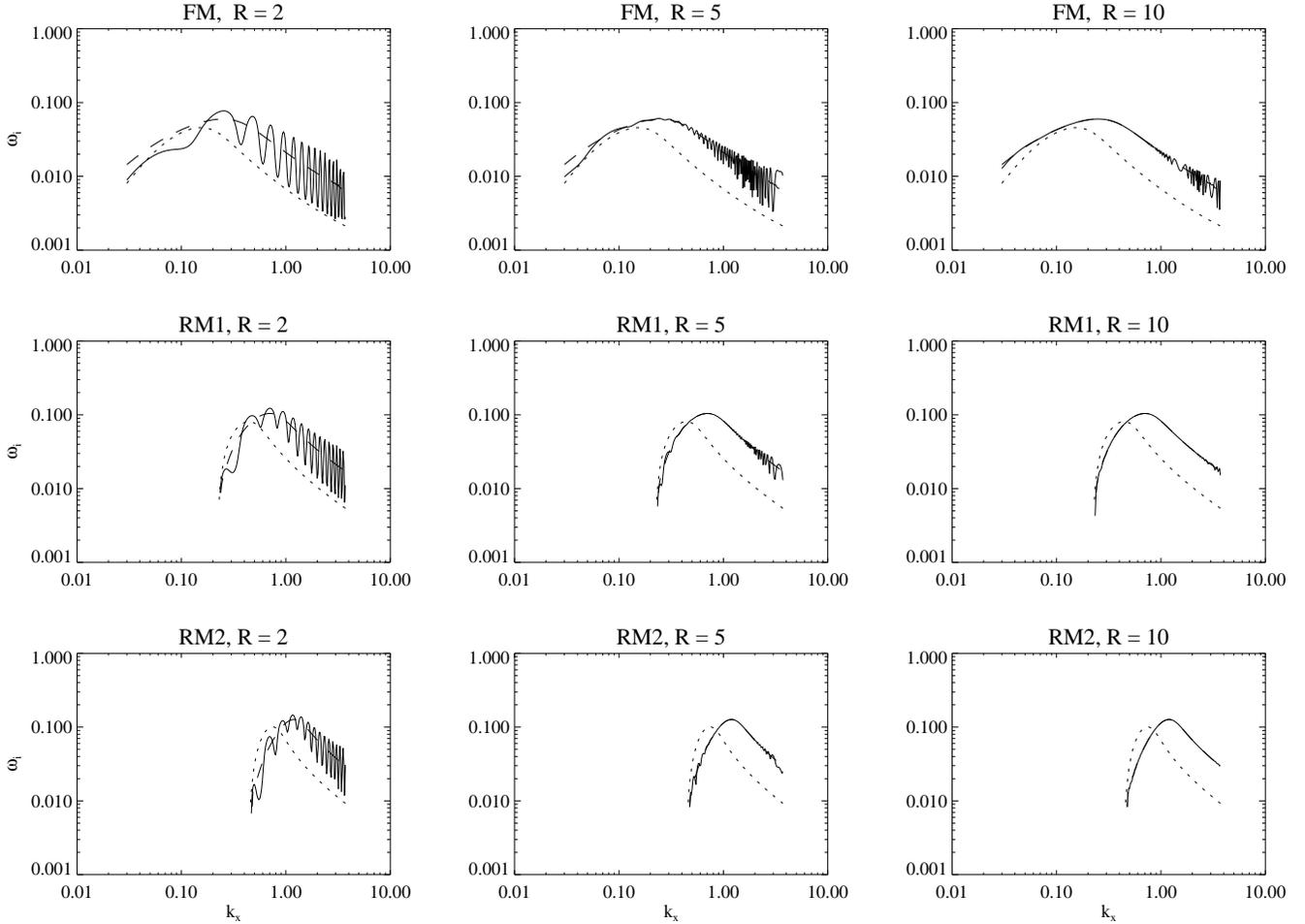}
\caption[]{
Temporal growth rate of the antisymmetric
solution for an underdense core and sheet jet, for fundamental mode
(top), first reflection mode (middle) and second reflection mode
(bottom).  $R=R_s/R_c =2$, $5$ and $10$ for first, second and third
columns.  Dotted lines correspond to a single jet case with no sheet and
$R=1$, while dashed lines correspond to a single jet case with no
external component and $R=\infty$ (infinite sheet).  Parameters are
$\gamma =10$, $M_s=0$, $\nu_c=\rho_c /\rho_s =0.01$ and
$\nu_s=\rho_s /\rho_e =0.1$
}
\end{figure*}

\begin{figure*}      
\epsfxsize=\hsize \epsfbox{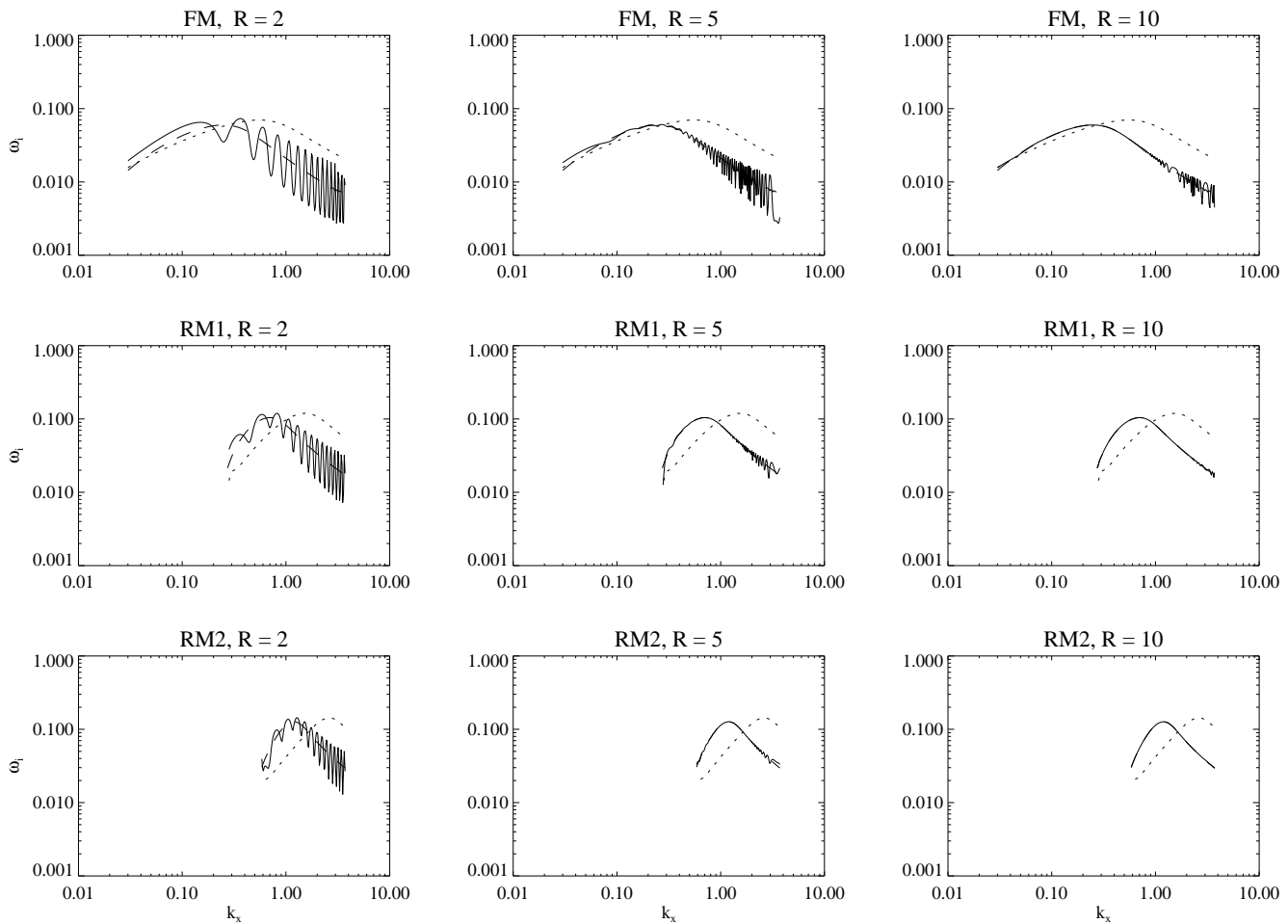}
\caption[]{
Same as  Fig. 2 but here $\nu_c=0.01$ and $\nu_s = 10$
}
\end{figure*}

\begin{figure*}        
\epsfxsize=\hsize \epsfbox{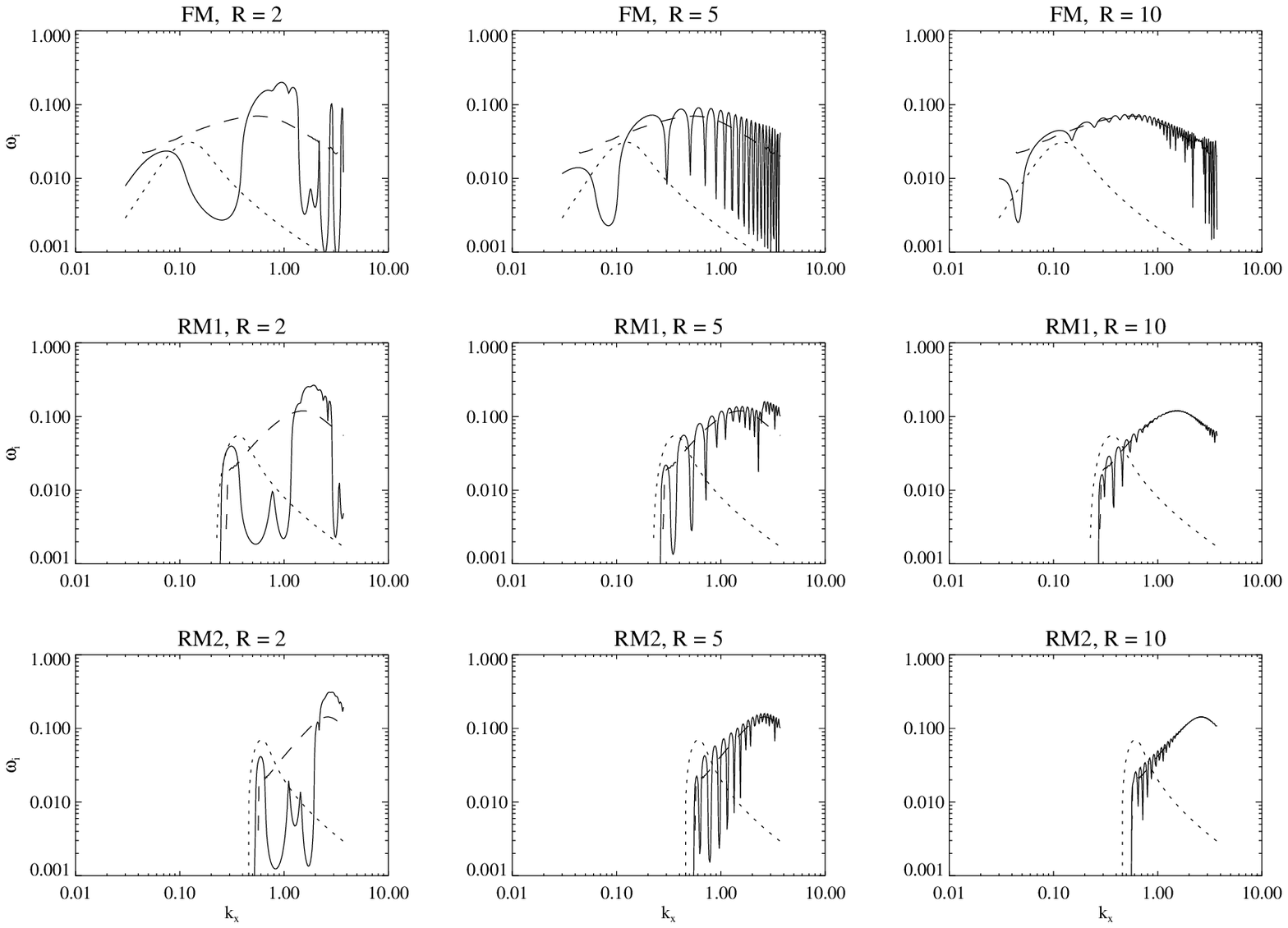}
\caption[]{
Same as  Fig.~2 and 3 but here $\nu_c=0.1$ and $\nu_s = 0.001$
}
\end{figure*}

We can take the forms of the transversal displacements as follows
\begin{equation} h'_{c,s} = A_{c,s} \exp \, i \, (k_x x - \omega t) + c.c.
\end{equation}
After substitutions and elimination of $A_{c,s}$, we reach the general
dispersion relation for core-sheet jets
\begin{equation}
 \left({Z_s \over Z_c}\right)\ {F^+_s(1) + {\cal R}_{se} F^+_s (2R-1)\over
   F^+_s(1) - {\cal R}_{se} F^+_s (2R-1)} =
  {F^+_c(1) + \varepsilon F^-_c(1) \over F^+_c(1) - \varepsilon F^-_c(1)}
   \label{eq:32}
\end{equation}
which leads to
\begin{equation}
{Z_s\over Z_c} \left[ {1+{\cal R}_{se} e^{2ik_{s_z} (R-1)} \over
 1-{\cal R}_{se} e^{2ik_{s_z} (R-1)} } \right] =
 \left\{\matrix{ {\rm coth}\ ik_{c_z}\quad {\rm for}\ \varepsilon =1\cr
 {\rm or}  \cr  {\rm th}\ ik_{c_z}\qquad {\rm for}\ \varepsilon =-1\cr }
  \right. \label{eq:33}
\end{equation}
where we have used the fact that by definition $F^+_s (R) F^-_s (1)
/ F^-_s (R) = F^+_s (2R-1)$ and have imposed the conditions of symmetry
or antisymmetry of the solutions, $G^-_c = \varepsilon G^+_c$ with
$\varepsilon = \pm 1$.
We introduce the complex normal acoustic
impedances (Payne \& Cohn 1985) for the core, sheet and external medium, which
after elimination of factors constant across all the components can be
expressed as
\begin{eqnarray}
Z_c &=& \Gamma_c \omega^2_{c_0}/k_{c_z},\\
Z_s &=& \omega^2_{s_0}/(\nu_s k_{s_z}),\\
Z_e &=& \omega^2/(\nu_s\nu_e k_{e_z}).
\end{eqnarray}
The reflection and transmission coefficients at core-sheet and sheet-external gas
interfaces are respectively defined as
\begin{eqnarray}
{\cal R}_{cs} = \frac{Z_s -Z_c}{Z_s +Z_c},&&
{\cal T}_{cs} = \frac{2 Z_s}{Z_s+Z_c} \label{RT_sc}\\
{\cal R}_{se} = \frac{Z_e -Z_s}{Z_e +Z_s},&&
{\cal T}_{se} = \frac{2 Z_e}{Z_e+Z_s} \label{RT_se}
\end{eqnarray}

When there is no reflection at this external boundary (interface
sheet-external), ${\cal R}_{se}=0$ and the effect of the external gas
disappears. Equations (\ref{eq:33}) reduce to the case of a single jet with
internal medium corresponding to the core component and external medium
to the sheet component.  Another way to cancel the effect of external
gas is to assume a very large sheet since the factor of ${\cal R}_{se}$ in
(\ref{eq:33}) vanishes as $\exp [-2R\, \Im(k_{s_z})]$ for $R>> 1$.  A
stratified jet with infinite sheet therefore corresponds to a single jet
with only core and sheet components. Conversely, for $R=1$, internal
and external interfaces coincide and equations (\ref{eq:33}) reduce
again to the case of a single jet, but then with only core component and
external gas.

\section{Numerical results and their physical interpretation}

The method used to solve the dispersion relation (\ref{eq:33}) is based
on the Newton-Raphson method (Press et al, 1992).
Different sets of parameters have been investigated.  They all lead to
the same qualitative results.
\begin{enumerate}
\item The presence of the sheet induces an
oscillating pattern on  the growth rate diagram
\item The stability properties
are dominated by the sheet even if its thickness is comparable with the core
radius.
\item The amplitude of oscillations of growth rate diminish with increasing
sheet thickness.
\end{enumerate}
This is well illustrated in  Figs.~2 to 4 which show the temporal growth rate
of the antisymmetric fundamental, first and second reflection modes of
core and sheet jets.

The apparent oscillating pattern is a new feature
caused by the reflections of acoustic waves at the interface between
sheet and external gas. These reflections are
represented by the reflection coefficient ${\cal R}_{se}$ in dispersion relation
(\ref{eq:33}). The wave radiated at the core-sheet ($cs$) interface toward
the sheet layer is reflected
from the sheet-external ($se$) interface and next returns to the ($cs$)
interface as shown in Fig.~5.
Thus, one can expect resonances due to the presence of the sheet component if
\begin{equation}
\Delta \phi =2 n \pi,  \label{rescond0}
\end{equation}
where $\Delta \phi$ is the change of phase of sound wave in sheet on the path
between $cs$, $se$ and $cs$ interfaces and $n$ is an integer number.
Introducing the length of path of the wave between $cs$ and $se$ interfaces
(see Fig.~5)
\begin{equation}
L=\frac{R-1}{\cos \alpha},
\end{equation}
where $\alpha$ is the wave propagation angle defined by $\ tg\alpha =k_x
/k_{s_z}$, and the wavelength
\begin{equation}
\lambda_s = 2\pi/ \sqrt{k^2_x + k^2_{s_z}}
\end{equation}
we can write the resonant condition (\ref{rescond0}) in explicit form
\begin{equation}
2 L = n \lambda_s \ ,\label{rescond1}
\end{equation}
The above interpretation of the oscillating pattern is demonstrated in Fig.~6.,
where we presented again the solution of Fig.~4 in the linear scale.
The superimposed high bars represent these values of $k_x$ (one should
remember that $k_{s_z}$ is a function
of $k_x$) for which the condition (\ref{rescond1}) is fulfilled.
For comparison, the  small bars represent these $k_x$ for which
we have an antiresonance
\begin{equation}
2 L = \left(n+\frac{1}{2}\right) \lambda_s \ ,\label{antirescond}
\end{equation}
We can easily notice a coincidence of the maxima of growth rate with
high bars and minima with small bars, what proves the validity of our
physical interpretation.  In a case of external gas which is significantly
lighter then the sheet gas we notice ${\cal R}_{se} \simeq -1$ (the change
of phase by $\pi$) resulting in a reversed relation between bars and the
maxima of growth rate. This effect is remarkable comparing Figs.~2 and 3,
where maxima and minima in both the figures are inverted.
 Figure ~5 shows a clear coincidence of maxima, minima and bars
described above. Varying the parameters of the core sheet configuration we
observe however that the bars can be shifted uniformly with respect to maxima
and minima toward smaller or larger $k_x$ values. This is not a surprise since
all the reflection and transmission coefficients are complex quantities in
general and the additional phase shifts are natural for each reflection at
the interfaces.

The entire interaction of sound waves at the core-sheet interface appears to
be very complicated.  The normal acoustic impedances $Z_c$ and $Z_s$ depend on
both real and oscillating imaginary parts of $\omega$ and $k_{c,s_z}$.  As Payne
\& Cohn (1985) shown the instability coincides with vanishing (in practice
small) values of denominator in expression for the reflection and transmission
coefficients.  In the case of core-sheet interface we have $Z_c \simeq -Z_s$
and even relatively small oscillations of these quantities result in strong
oscillations of ${\cal R}_{cs}$ and ${\cal T}_{cs}$ around the averaged values
typical for the sheet layer of infinite thickness.  Both the reflected and
transmitted waves at the ($cs$) interface come back to this interface after a
finite number of cycles returning the acoustic energy which has already been
radiated in both directions.  This returned energy results in the modified
growth rate of the instability and again the growth rate influence the
reflection and transmission coefficients at the ($cs$) interface.
Because of the multiplicity of interfaces, many waves interfere within the
core-sheet jet configuration and the detailed description of these
interactions seems to be a difficult task.

\begin{figure}         
\epsfxsize=\hsize \epsfbox{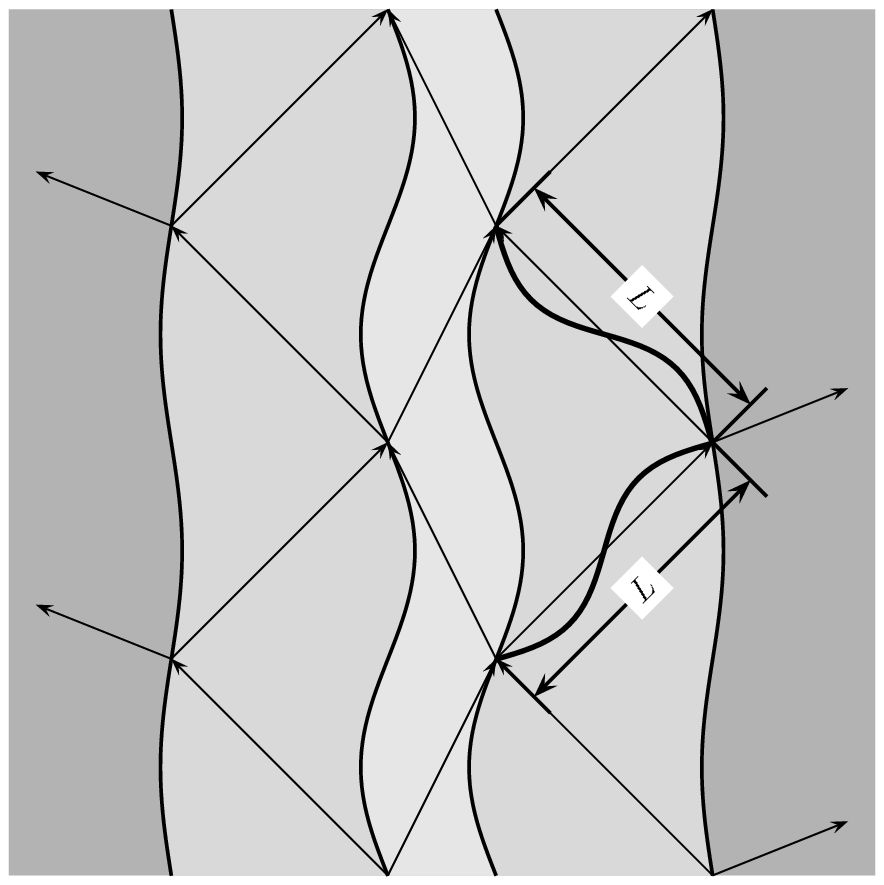}
\caption[]{
The physical interpretation of the effect of oscillating growth rate is as
follows: The sound wave emitted at the internal interface first reflects
at the external interface and then returns to the internal interface and
interferes with the new emitted wave.  The waves drawn within the sheet
layer demonstrate
the integer number of wave periods between interfaces, not the way of
deformation of sheet (the acoustic waves are longitudinal waves).  The case
$L=\lambda_s$ is shown.
}
\end{figure}

\begin{figure}          
\epsfxsize=\hsize \epsfbox{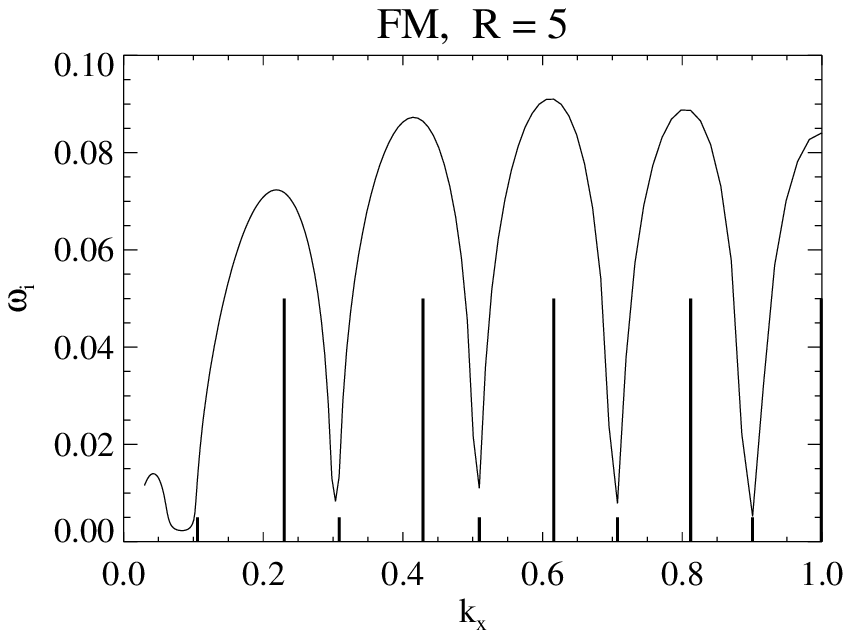}
\caption[]{
The maxima of the growth rate coincide with the integer multiples
of wavelength on the path of the acoustic wave. We show the growth rate
versus wavenumber k in the linear scale for the case presented in
Fig.~4  (fundamental mode, R = 5).  The high bars
located below the curve of growth rate indicate the fulfillment of the
resonance condition (\ref{rescond1}) and the small bars indicate the
fulfillment of the antiresonance condition  (\ref{antirescond})
}
\end{figure}

We explored different sets of values for the densities, Lorentz factor of the
core and Mach number of the sheet, for antisymmetric or symmetric solutions.
In all chosen cases, $|{\cal R}_{se}|$ remains smaller than 1, implying
that no magnification of the waves occurs at the external interface.
Variation of the sheet Mach number slightly shifts the oscillating pattern
but does not modify the average growth rate as long as the sheet keeps
a classical velocity. This statement is valid also in the case of sheet
velocity exceeding the sheet sound speed ($M_s>1$) as long as we
assume that the core is relativistic and sheet is nonrelativistic.
The main source of the
instability comes from the internal interface where amplification of the
waves can take place. Generally speaking, the instability develops due to
the interaction between the core and the sheet, with the
oscillatory modification superimposed by the presence of the external
($se$) interface.

For small $k_{s_z}$, when the effective thickness of the sheet is
smaller than the half-wavelength, namely $\lambda_s >2(R-1)/\cos \alpha$ the
growth rate for the stratified case resembles to the solution obtained for a
single jet without sheet $(R=1)$ or appears as a kind of compromise between the
two extreme cases of no sheet or infinite sheet.

For all studied sets of parameters, the temporal growth rate of the stratified
jet instability tends to the solution obtained for the case of an infinite
sheet or oscillates around it for large $k_{s_z}$ (corresponding to large
$k_x$).
Oscillations are always reduced while increasing the sheet
thickness.  Practically all solutions that we obtained converge to the case of
infinite sheet as soon as $R$ reaches the order of 5 and often even less.  So
the presence of a sheet with a thickness of only a few times the core component
radius imposes the growth rate of the instability.  It is the physical
parameters of the sheet relatively to the core medium which are decisive for
the growth of the instability instead of the parameters of the external medium.
An illustration of this is provided by Fig.~2 and Fig.~3 which show almost
similar growth rate for the stratified scenario (except for the inversion of
the maxima and minima) in spite of different values of the density ratio
$\nu_e=\rho_s /\rho_e$.

\section{Astrophysical implications}

The effect of the sheet that we study in the 2-dimensional slab geometry
corresponds to the effect due to an envelope, a sheath or a cocoon around the
jet in the cylindrical description. The general result that properties of the
sheet essentially influence the linear growth of the Kelvin-Helmholtz
instability have several consequences.

It shows that the maximal growth rate and corresponding wavenumber can
be modified by the presence of even a modest envelope. This is illustrated
by the case of Fig.~4, where an envelope not very different from the
core-component (same thickness with $R=2$, density ratio of $\rho_c
/\rho_s =0.1$) induces an increase of $\omega_I^{\max}$ from 0.03 to 0.2
and of $k^{\max}_x$ from 0.1 to 1 for the fundamental mode.
The linear growth of the helical mode of such a jet is therefore much
faster and the resonant wavelength decreases to about $6R_c$.  Such
modification due to the new complexity driven by the envelope-effect can
help to solve some difficulties encountered when trying to deduce
parameters such as densities and temperatures from the radio morphology
of jets, under the assumption that knots and wiggles are respectively
due to the growth of pinching and helical modes (Ferrari et al, 1983;
Zaninetti, Van Horn, 1988).  Indeed, the parameters deduced in such
studies for the external medium can be considered as applying only to an
envelope surrounding the jet, with density and temperature possibly
quite different from values deduced for instance from X-rays data on
intergalactic gas. However the model of
core-sheet jet is complex and involves much more free parameters
which makes any model-dependent predictions more difficult.

The determination of physical parameters of astrophysical jets based on the
model of Kelvin -- Helmholtz instability is a challenge, but it is also made
difficult by our lack of detailed knowledge on the nonlinear evolution of the
instability.  Several authors investigated the nonlinear regime by means
of numerical simulations (e.g.  Hardee et al.  1992, Bodo et al.  1994)
in the case of nonrelativistic jets and applying analytical methods
(Hanasz 1995) in the case of relativistic jets.
The last approach shows that in the phase preceding the formation of shock
waves within the relativistic jet, the growth of instability slows down while
the perturbation amplitude grows and finally the growth is stopped for the
amplitude of lateral displacement of the jet which is only a fraction of the
jet radius.  The physical reasons responsible for this behaviour can be
summarized as a strong nonlinear growth of modulus of the acoustic impedances
of both the internal (relativistic) and the surrounding (nonrelativistic)
gases, associated with the growth of perturbation amplitude.  The properties of
the surrounding gas and the presence or absence of the sheet layer are
important for the properties of the Kelvin-Helmholtz instability in this aspect
as well.  The numerical simulations allow to trace the evolution of
perturbations up to the jet disruption.  As Bodo et al.  (1994) shown the time
interval between the formation of first shock waves and the beginning of the
mixing phase is dependent on the values of the Mach number of the jet and the
density contrast.  According to this work the smaller density and higher Mach
number jets can survive in the form of laminar flow for a longer time.  We
would like to point out that in analogy to our previous results we can expect
that the parameters of sheet in the case of stratified jets will control the
life time of the laminar flow.

Coming back to the linear approximation we point out that the presence of an
envelope underdense relatively to the external medium tends to induce faster
growth of the instability while overdense envelope slows it down.  The high
stability of a large number of radio jets can therefore be better understood if
dense and cool envelopes are commonly present, as expected in ``multilevel'' or
two-component jet models (Smith, Raine, 1985; Baker et al, 1988; Sol et al,
1989; Achatz et al, 1990).  This could explain for instance the remarkable
stability of the inner jet of the wide-angle tail (WAT) radio source 3C465.
This source is hosted by the central galaxy of the cluster of galaxies Abell
2634.  A hot intercluster gas makes the external medium in which the radio jet
propagates.  X-ray observations provided estimates of its temperature and
density with distance to the core, $T_e \simeq 2\times 10^7$K and $n_e \simeq
8\times 10^{-4} (p^+e^-)$~cm$^{-3}$ at 20~kpc (Eilek et al, 1984).  From
Faraday rotation measures, Leahy (1984) suggested the existence of thermal
plasma inside the jet with a density of $10^{-2} (p^+e^-)$~cm$^{-3}$ in the
inner hot spots at about 20~kpc from the core and of $5\times 10^{-4}
(p^+e^-)$~cm$^{-3}$ on average in the tails.  We tentatively identify this
thermal plasma with the sheet component of our study (indeed in our
description, the slow plasma of the sheet can also be present at $z<R_c$, with
the fast-moving plasma of the core just streaming through it).  In two-flow
models, the slow component (here the sheet or envelope) is made by a collimated
wind coming from the accretion disc.  Assuming that its temperature is
comparable to or less than the one of the gas in the disc at the basis of the
outflow, one can expect a temperature in the range $10^4 -10^6$K for the
envelope, from the temperature variation proposed by Phinney (1989) in the
accretion discs.  Taking $T_s\lta 10^6$ required $n_s \gta 1.6\times
10^{-2}(p^+e^-)$cm$^{-3}$ in the region of the inner hot spots to ensure
pressure equilibrium with the intercluster gas at the external interface.  Such
values of $n_s$ are quite in agreement with the Faraday rotation analysis and
correspond to an overdense envelope which tends to stabilize the inner jet
structure.

Such type of envelopes
could be present in many other extragalactic radio sources as proposed in an
application of two-component jet models (Dole et al, 1995) and as
suggested by the ``tomography'' analysis of radio data (Rudnick, 1995).
Direct estimates of the envelope densities are not yet available. However
the sheath detected by Katz-Stone and Rudnick (1994) around the jet of
the FRII source Cygnus A corresponds to an enhancement of the density
of radiating particles, and therefore likely to an increase of the
thermal plasma density as well. Overdense sheaths would influence the
dynamics of FRII jets and increase their stability. A somewhat analogous
role was already known to be played by FRII cocoons. Indeed numerical
simulations emphasized the importance of a cocoon for the
dynamics of FRII jets (Norman et al, 1984a,b) and identified two regimes,
mode-dominated or cocoon-dominated. Our study deal with laminar flows
and can not precisely describe FRII-cocoons which are turbulent. However
it proposes some theoretical interpretation of the cocoon-dominated
regime observed in numerical experiments. Our results are analogous to the
results obtained by Hardee \& Norman (1990), who notice that the jet
dynamics is determined by properties of the jet and lobe, but not by the
properties in the undisturbed medium. In their numerical simulations the
lobe surrounds the jet in a manner resembling to our sheet or envelope.
In fact, four different layers can be expected in such cases, namely the
inner jet, the sheath, the cocoon and the external media. From our
results on core-sheet jets, we tentatively infer that the stability
properties of this highly stratified configuration will be dominated by
the Kelvin-Helmholtz instability of the most unstable interface, likely
the internal one if the inner jet is relativistic, slightly modified by
the multiple reflections of the acoustic waves at the other interfaces.

The question of the origin of sheaths or envelopes around extragalactic
jets is still open. They can just emanate from the inner jets as the
result of particle diffusion. This could concern the slow-moving radio
``cocoon'' of the jet in 3C273. The interaction itself between the jet
and its surroundings induces turbulent sheared and entrainment layers
which can lead to the formation of sheaths for instance in FRI sources.
In FRII sources, the backflow from the hot spots constitutes some kind of
sheaths as well,  as already mentioned. Other possibilities proposed by
two-component models for jets are to generate stratified jets, inner
cores and surrounding sheaths, directly from the central engines.
In all these cases,
the presence of an envelope with thickness comparable or slightly larger
than the central jet radius deeply modifies the interaction of the jet
with the ambient medium of the host galaxy, and somewhat isolates the
jet from its surroundings. This makes then rather problematic to explain
the dichotomy between radio-loud and radio-quiet objects by the
difference in the media where the jets have to propagate. Such an
argument of difference in the host galaxy interstellar medium is often
proposed to account for the fact that well-developed radio jets
are observed in elliptical galaxies but not in spiral and Seyfert
galaxies, based on the idea that jets are rapidly destroyed by
their interaction with the dense and inhomogeneous interstellar gas
expected in spiral galaxies. In fact, the interstellar medium can
influence the radio source evolution in two ways, first by its direct
interaction with the jet, second by its primordial role during the
accretion process. Our results favour the latter option since the former
one appears easily inhibited by specific internal jet properties.
Within such a view, the dichotomy between radio-loud and radio-quiet
objects should arise from specific properties of the central engine
possibly acquired during its formation and accretion phase.

\begin{acknowledgements}
This project was partially supported by PICS/CNRS no.198
``Astronomie Pologne''.
\end{acknowledgements}


\begin{thebibliography}{}
\bibitem{} Achatz, U., Lesch, H., Schlickeiser, R., 1990, \aaa, 233, 391.
\bibitem{} Achatz, U., Schlickeiser, R., 1992, {\it Extragalactic radio sources
           --~from beams to jets}, Roland, J., Sol, H., Pelletier, G., ed.,
           Cambridge University Press, p.~256.
\bibitem{} Bahcall, J.N., Kirhakos, S., Schneider, D.P., Davis, R.J.,
           Muxlow, T.W.B., Garrington, S.T., Conway, R.G., Unwin, S.C., 1995,
           \apj, 452, L91.
\bibitem{} Baker, D.N., Borovsky, J.E., Benford, G., Eilek, J.A., 1988,
           \apj, 326, 110.
\bibitem{} Biretta, J.A., Zhou, F., Owen F.N., 1995, \apj, 447, 582.
\bibitem{} Birkinshaw, M., 1991, {\it Beams and jets in Astrophysics}, P.A.
           Hughes, ed., Cambridge University Press, p.~278.
\bibitem{} Bodo, G., Massaglia, S., Ferrari, A., Trussoni, E., 1994, \aaa,
           283, 655
\bibitem{} Chan, K.L., Henriksen, R.N., 1980, \apj, 241, 534.
\bibitem{} Dole, H., Sol, H., Vicente, L., 1995, IAU Symposium 175
           `Extragalactic Radio Sources' (Bologna).
\bibitem{} Eilek, J.A., Burns, J.O., O'Dea, C.P., Owen, F.N., 1984, \apj,
           278, 37.
\bibitem{} Ferrari, A., Trussoni, E., Zaninetti,L., 1978, \aaa, 64, 43
\bibitem{} Ferrari, A., et al, 1982, \mnras, 198, 1065.
\bibitem{} Ferrari, A., et al, 1983, \aaa, 125, 179.
\bibitem{} Hanasz, M., 1995, PhD Thesis, Nicolaus Copernicus University, Torun.
\bibitem{} Hardee,  1986, \apj, 303, 111.
\bibitem{} Hardee, P.E.,  Norman, M.L., 1988, \apj, 334, 70
\bibitem{} Hardee, P.E., Norman, M.L., 1990, \apj, 365, 134
\bibitem{} Hardee, P.E., Cooper, M.A., Norman, M.L., Stone, J.M. 1992,
           \apj, 399, 478
\bibitem{} Katz-Stone, D.M., Rudnick, L., Anderson, M.C., 1993,
           \apj, 407, 549.
\bibitem{} Katz-Stone, D.M., Rudnick, L., 1994, \apj,
           426, 116.
\bibitem{} K\"onigl, A., Kartje, J.F., 1994, \apj, 434, 446.
\bibitem{} Landau, L.D., Lifshitz, E.M., 1959, {\it Fluid Mechanics},
           Pergamon Press, Oxford.
\bibitem{} Leahy, J.P., 1984, \mnras, 208, 323.
\bibitem{} Melia, F., K\"onigl, A., 1989, \apj, 340, 162.
\bibitem{} Norman, M.L., Smarr, L.L. and Winkler, K.-H.A., 1984a in {\it
           Numerical Astrophysics: A Festschrift in Honor of James R. Wilson},
           ed. R. Bowers, J. Centrella, J. Leblanc and M. Leblanc
           (Jones and Bartlett: Boston).
\bibitem{} Norman, M.L., Winkler, K.-H.A., Smarr, L.L., 1984b in {\it Physics
           of Energy Transport in Extragalactic Radio Sources}, eds A.H. Bridle
           and J.A.Eilek, NRAO, p. 150
\bibitem{} Norman, M.L.,  Hardee, P.E.,   1988, \apj, 334, 80.
\bibitem{} O'Donoghue, A.A., Owen, F.N., Eilek, J.A., 1990, \apj, 72, 75.
\bibitem{} O'Donoghue, A.A., Owen, F.N., Eilek, J.A., 1993, \apj, 408, 428.
\bibitem{} Payne, D.G., Cohn, H., 1985, \apj, 291, 655.
\bibitem{} Pelletier, G., Sol, H., 1992, \mnras, 254, 635.
\bibitem{} Phinney, E.S., 1989, {\it Theory of accretion disks}, Meyer, F.,
           Duschl, W.J., Frank, J., Meyer-Hofmeister, E., ed.,  Kluwer Academic
           Publishers, p.~457.
\bibitem{} Press, W.H., Flannery, B.P., Teukolsky, S.A., Vetterling, W.T.,
           1992, Numerical recipes, Cambridge University Press.
\bibitem{} Rudnick, L., Katz-Stone, D.M., Anderson, M.C., 1994, \apj
           Suppl. Ser., 90, 955.
\bibitem{} Rudnick, L., 1995, IAU Symposium 175 `Extragalactic Radio Sources'
           (Bologna).
\bibitem{} Smith, M.D., Raine, D.J., 1985, \mnras, 212, 425.
\bibitem{} Sol, H., Pelletier, G., Ass\'eo, E., 1989, \mnras, 237, 411.
\bibitem{} Taub, A.H., 1948, Phys. Rev. 74, 328
\bibitem{} Zaninetti,   Van Horn,    1988, \aaa, 189, 45.
\end{thebibliography}
\end{document}